# Using a remote control to determine the infrared absorption coefficient in water

Santiago Ortuño-Molina, Adrián Garmendía-Martínez, Francisco M. Muñoz-Pérez, Juan C. Castro-Palacio, and Juan A. Monsoriu

*Centro de Tecnologías Físicas, Universitat Politècnica de València, Camino de Vera, s/n, 46022 València, Spain.*

## Abstract:

In this work, we present a simple and low-cost experiment designed to determine the infrared water absorption coefficient. We used a TV remote control as a point source of infrared light. The intensity after passing the light through different heights of a water column is measured with a solar cell connected to a speaker. The recorded signal, captured with a smartphone sound recorder, provides a practical demonstration of the Beer-Lambert law. The collected data were fitted to the theoretical model, obtaining a very good agreement between the value of the water absorption coefficient obtained experimentally and that reported in the literature.

## 1. Introduction

The study of the properties of light and the phenomena that occur when it interacts with matter has captivated the scientific community. This is also a subject that is taught in high school and university levels. One such phenomenon is the absorption of light in a medium like water and its behavior as it passes through. It is widely known that the intensity of a wave decreases with the inverse square of the distance (Resnick *et al.*, 1997; Hecht, 2017). Known as the "inverse square law," this physical concept can be found in various natural phenomena such as gravitation, electrostatics, electromagnetic radiation, and sound waves (Gatzia & Ramsier, 2021; Williams *et al.*, 1971). However, while this law describes the attenuation of light intensity, when we seek to understand the attenuation of light within a solution, the Beer-Lambert law is the appropriate one. The Beer-Lambert law is a powerful tool in the quantitative analysis of light attenuation in solutions, relating absorbance to concentration and the optical path length in materials different (Holler & Skoog, 2017; Colt *et al.*, 2020). Teaching the Beer-Lambert law provides undergraduate students with a fundamental understanding of the characteristics of light and its interaction with a medium.

On the other hand, the incorporation of new technologies in the teaching of physical concepts has gained relevance, as it allows the creation of new tools in the students' learning process. One such technology is the smartphone, which, with its wide range of sensors, can generate experiences as an alternative to traditional practices. Currently, various studies in the literature incorporate smartphones (Monteiro & Martí, 2022; Torriente-García *et al.*, 2023b; Torriente-García *et al.*, 2024). An example of this is the use of the ambient light sensor to determine the efficiency of light sources, teach the inverse square law, or study oscillations (Sans *et al.*, 2013; Sans *et al.*, 2017). A previous work has shown that combining technologies such as small solar cells and the sound recorder of a smartphone provides an alternative way to teach the inverse square law using the infrared signal from a remote control (Marín-Sepúlveda, 2024). Additionally, some educational studies address the absorption of ultraviolet and infrared light by glass, as well as the Beer-Lambert law to determine the absorption coefficient of the intensity of visible light through different materials by changing the molar concentration of translucent chemical solutions (Christopher et al., 2024; Chen *et al.*, 2024; Onorato et al., 2018).

In this work, we present a simple and low-cost experiment to determine the water absorption coefficient for infrared light. A TV remote control was used as a point source of infrared light. The intensity was measured after the light beam passes through different heights of a water column by a solar cell connected to a loudspeaker. The resulting signal was registered with a smartphone sound recorder. This experiment aims to provide a practical application of the Beer-Lambert law, demonstrating its principles in an accessible and innovative way. By integrating common technologies such as smartphones, this work offers an alternative approach to traditional laboratories, enhancing the educational experience for students.

## 2. Experimental setup and physical model

The Beer-Lambert law is a relationship between the attenuation of light through a medium and the properties of that medium. Mathematically, the intensity of monochromatic light transmitted through a medium is given by

$$I(x) = I_0 \cdot e^{-\alpha x} \qquad (1)$$

where $I_0$ is incident intensity of light, $I(x)$ is the intensity after it has travelled a distance $x$ in the medium, and a is the absorption coefficient. In this work, the absorption of infrared radiation was measured as a function of the thickness of a column of water. When a light wave hits a transparent or translucent medium, part of the incoming energy ($I_0$) is transmitted ($I$). The quotient $\frac{I}{I_0}$ is known as transmittance $T$. Figure 1 includes a visual representation of this phenomenon.

This exponential behavior of the transmitted intensity can be experimentally verified by using the IR signal from a remote control working as a point source. The signal intensity is measured with a solar cell connected to a speaker, which produces a sound when it receives the electric signal provided by the solar cell.

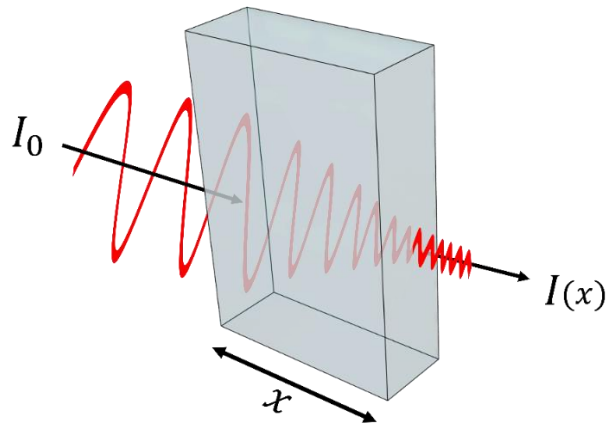

**Figure 1.** Transmittance of a light wave through an absorbent material of thickness **x**.

The sound is then recorded with a smartphone, and the intensity (in arbitrary units) is obtained by editing the audio file.

Figure 2 shows the experimental setup used in this work to study the absorption of infrared radiation in water. The TV remote control is attached to a support stand and placed above the water column. This remote control contains an infrared LED of $l \sim 929$ nm.

A narrow glass container is placed over a single solar cell. For different heights of the water column in the cylindric container, the remote-control play button is pressed three times, and the corresponding sound emitted by the loudspeakers connected to the cell is recorded with a smartphone. Each pulse of the remote control contains a changing intensity encoding a set of audible frequencies which are reproduced at the speaker.

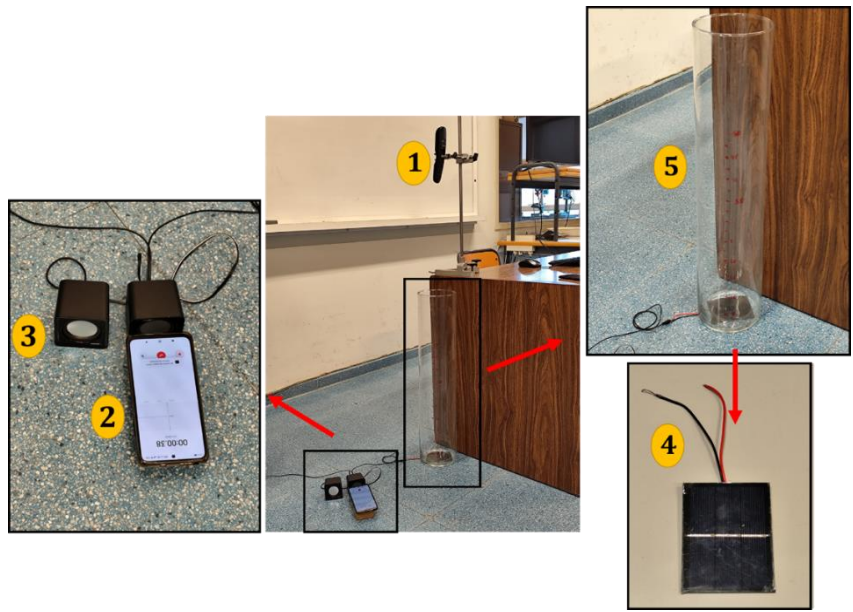

**Figure 2**. Experimental setup used to measure the absorption of the infrared signal in a column of water. It includes a remote control (1), a smartphone (2), speakers (3), and a solar cell (4) placed under the cylindrical vase (5) that contains water

## 3. Results and discussion

Figure 3 presents the edited view of the digital audio file recorded with the smartphone for different heights of the water column in the cylindrical glass container. The height of the water column is increased by adding equal amounts of water every time. The peaks represent the three pulses resulting from pressing the power button on the remote control three times. In Table 1, the arithmetic mean of these three recorded pulses is shown for different heights of the water (numeric labels in Fig. 3). Measurements were taken at 2.5 cm intervals. The distance between the remote control and the solar cell was kept constant throughout the experiment. The figure illustrates the decreasing variation in audio amplitude with increasing water column height. The table starts at 5 cm because, below this distance, the signal is not clearly detected.

**Table 1**. Experimental data points.

| # | Thickness $x$ (cm) | Intensity $I$ (a.u.) | ln ($I$) |
|---|---|---|---|
| 1 | 5.0 | 4449 | 8.40 |
| 2 | 7.5 | 3035 | 8.02 |
| 3 | 10.0 | 2077 | 7.64 |
| 4 | 12.5 | 1704 | 7.44 |
| 5 | 15.0 | 1378. | 7.23 |
| 6 | 17.5 | 1106 | 7.01 |
| 7 | 20.0 | 790 | 6.67 |
| 8 | 22.5 | 553 | 6.32 |
| 9 | 25.0 | 414 | 6.03 |
| 10 | 27.5 | 303 | 5.72 |
| 11 | 30.0 | 249 | 5.52 |
| 12 | 32.5 | 201 | 5.30 |
| 13 | 35.0 | 134 | 4.90 |
| 14 | 37.5 | 111 | 4.71 |

We have also included in Table 1 the logarithm of the registered intensities. Figure 4 shows the plot of $ln(I)$ as a function of the height of the water column $x$. From Eq. (1), a linear fit $ln(I) = -\alpha x + ln(I_0)$ was performed, obtaining an $\alpha = (0.113 \pm 0.002)$ cm$^{-1}$ and $ln(I_0) = 8.87$ with a correlation coefficient of 0.9969. If we measure the light intensity with the empty column ($x = 0$) we obtain $I_0 = 7480$, so $ln(I_0) = 8.92$. The discrepancy between this value and the corresponding are obtained with the fitted experimental data is lower than 1%. We can also compare the $\alpha$ value obtained from the fitting with that reported in the literature which is $\alpha = 0.119$ cm$^{-1}$ (Kou *et al.*, 1993), for the λ used, obtaining a good agreement between both results.

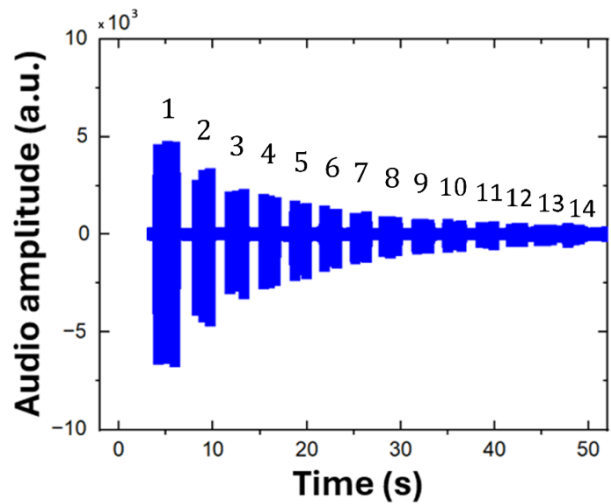

**Figure 3**. Edited version of the audio file containing the sound produced by the infrared signal incident on the solar cell. It shows the variation in the intensity of the infrared radiation with the height of the water column. In this case, the distances between consecutive peaks correspond to 2.5 cm.

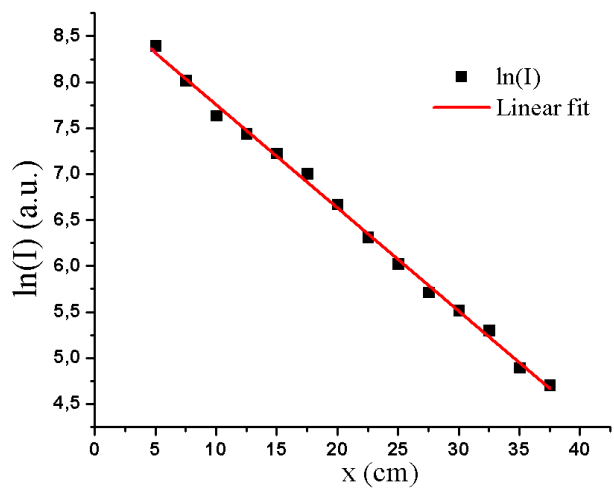

**Figure. 4.** Logarithm of the registered intensity as a function of the height of the water column.

## 4. Conclusions

The intensity of infrared radiation as a function of the height of a water column has been measured experimentally. The setup consists of a remote control, a cylindrical container with water, a solar cell connected to a speaker, and a smartphone with a sound recorder. The simplicity and low cost of this experimental setup make it easily implementable in high school and early university physics courses. From the experimental data, the absorption coefficient of infrared radiation in water has been determined. This value, when compared with the literature, shows a good agreement.

## Acknowledgment

This work was supported by the Spanish Ministerio de Ciencia e Innovación (grant PID2022- 142407NB-I00) and by Generalitat Valenciana (grant CIPROM/2022/30), Spain. The authors would like to thank the Instituto de Ciencias de la Educación (Institute of Education Sciences) at the Universitat Politècnica de València (UPV) for its support to the teaching innovation group MSEL.